\documentstyle[sprocl,epsfig]{article}

\def\Journal#1#2#3#4{{#1} {\bf #2}, #3 (#4)}

\def\NPB{{\em Nucl. Phys.} B}
\def\PLB{{\em Phys. Lett.}  B}

\def\PRD{{\em Phys. Rev.} D}

\def\EPG{{\em Eur.Phys.J.} C}

\def\be{\begin{equation}}
\def\ee{\end{equation}}
\def\bea{\begin{eqnarray}}
\def\eea{\end{eqnarray}}

\begin{document}

\title{DESCRIPTION OF THE PROTON STRUCTURE FUNCTION $F_2^p(x,Q^2)$ IN THE
FRAMEWORK OF EXTENDED REGGE - EIKONAL APPROACH}

\author{V. A. PETROV and A. V. PROKUDIN}

\address{
\vskip 0.2cm 
Institute For High Energy Physics,\\ 
142284 Protvino , Russia\\E-mail: petrov@mx.ihep.su \\
prokudin@th1.ihep.su}
\vskip 0.2cm




\maketitle\abstracts{The proton structure function $F_2^p(x,Q^2)$ is
described in the framework of the off-shell extention of the
Regge-eikonal approach which automatically takes into account off-shell
unitarity. We achieved a good quality of description for $x<10^{-2}$ and
we
argue that the data on $F_2^p(x,Q^2)$ measured at HERA can be fairly
described with classical universal Regge trajectories. No extra, ``hard''
trajectories of high intercept are needed for that. The $x-, Q^2-$ slopes
and the effective intercept are discussed as functions of $Q^2$ and $x$.}

\section{INTRODUCTION}
In the literature one can find a number of models describing the behaviour
of $F_2^p(x,Q^2)$in the framework of so called ``soft" pomeron~\cite{models}
or with help of ``hard" pomeron~\cite{models1}.
We add our arguments in favour of the ``soft" pomeron by introducing
a
model taking into account unitarity for the processes with virtual
particles. We outline some basic properties of the
model that will help to distinguish it from the others.

\section{OFF-SHELL EXTENTION OF THE REGGE-EIKONAL APPROACH}

Construction of the model starts from the unitarity condition:
$$
Im T(s,\vec b) = \vert T(s,\vec b)\vert^2 + \eta (s,\vec b)
$$
(here $T(s,\vec b)$ is the scattering amplitude in the impact
representation,
$\vec b$ is the impact parameter, $\eta (s,\vec b)$ stands for the
contribution of 
inelastic channels)
which in case of eikonal amplitude
\begin{equation}
T(s,\vec b)=\frac{e^{2i\delta (s,\vec b)}-1}{2i}
\label{eq:ampl}
\end{equation}
(here $T(s,\vec b)$ is the scattering amplitude, $\delta (s,\vec b)$ is the
eikonal function), may be rewritten as 

\begin{equation}
Im \delta (s,\vec b) \ge 0, \; s>s_{inel}
\label{eq:euc}
\end{equation}

We choose the following eikonal function in $t$-space (here $t$ is
transferred momentum)
\be
\hat \delta (s,t) = c \Big( \frac{s}{s_0} \Big)
^{\alpha(0)}e^{t\frac{\rho^2}{4}}
\label{eq:eikonalt}
\ee
where  
\be
\rho^2 = 4\alpha'(0) ln\frac{s}{s_0}+r^2
\ee
\noindent
is reffered to as``reggeon radius''.

It means that the eikonal function has a simple pole in $J$ plane and the
corresponding Regge trajectory can be written in the following form:
\be
\alpha(t) = \alpha(0) + \alpha '(0)t
\label{eq:rt}
\ee

In order to rewrite functions in $t$- and $b$-spaces one uses Fourier-Bessel
transformation:
\be
\begin{array}{r}
\hat f(t)= 4 \pi s\int_{0}^{\infty} db^2 J_0(b\sqrt{-t}) f(b) \\
\\
f(b)= \frac{1}{16 \pi s}\int_{-\infty}^{0} dt J_0(b\sqrt{-t}) \hat f(t) 
\end{array}
\label{eq:fb}
\ee

Making use of Eq.~(\ref{eq:fb}) we obtain the following $b$-representation
of the eikonal function:
\be
\delta (s,b) = \frac{c}{s_0}
\Big(\frac{s}{s_0}\Big)^{\alpha(0)-1}\frac{e^{-\frac{b^2}{\rho^2}}}{4\pi
\rho^2}
\label{eq:eikonalb}
\ee

We would like to emphasise that the ``pomeron'' in our model is the leading
pole of the eikonal function.

For cross-sections we use the following normalizations:
\be
\begin{array}{l}
\sigma_{tot} = \frac{1}{s} ImT(s,t=0) \\
\\
\sigma_{el} = 4 \pi \int_{0}^{\infty}db^2 \vert T(s,b) \vert^2 \\
\\
\frac{d\sigma}{dt} = \frac{\vert T(s,t) \vert ^2}{16\pi s^2}
\end{array}
\label{eq:norm}
\ee

The off-shell extention of our approach can be resolved from the following
consideration.
$T(s,t)$ can be  rewritten in the following way:
\be
\begin{array}{l}
T(q',p'\vert q,p)= \hat \delta(q',p'\vert q,p)+ i \int
d^3q''d^3p''d^3q'''d^3p'''(2\pi)^4\delta(q'+p'-q''-p'')\cdot \\
\cdot (2\pi)^4\delta(q'''+p'''-q-p) \delta(p',q'\vert
q'',p'')L(q'',p''\vert q''',p''')\delta(q''',p'''\vert q,p)\\
\end{array}
\label{eq:expand}
\ee
where (in case of identical particles of mass $m$)
\be
\begin{array}{l}

T(s,t)= T(q',p'\vert q,p)\bigg\vert_{q'^2=q^2=p'^2=p^2=m^2}\\
\hat \delta(s,t)= \hat
\delta(q',p'\vert q,p)\bigg\vert_{q'^2=q^2=p'^2=p^2=m^2} \\
\\
s=(p+q)^2=(p'+q')^2 \\
\\
t=(p-p')^2=(q-q')^2 \\
\\
d^3p = d\vec p/ (2\pi)^3 2 p_0 \equiv d^4p\Theta(p_0)\delta(p^2-m^2)\\
\\
\sum_{n=2}^{\infty}\frac{2(2i\delta(s,\vec
b))^{n-2}}{n!}\equiv L(s,\vec b)\\
\\
L(s,t) = 4s\int d^2b e^{i\vec k \vec b}L(s,\vec b)\\
\\
\end{array}
\label{eq:expand1}
\ee
Eq.(~\ref{eq:expand}) can be symbolically represented by the following
figure:

\begin{figure}[h]
\vskip -5cm
\hskip 0cm {\vbox to 50mm{\hbox to 50mm{\epsfxsize=120mm
\epsffile{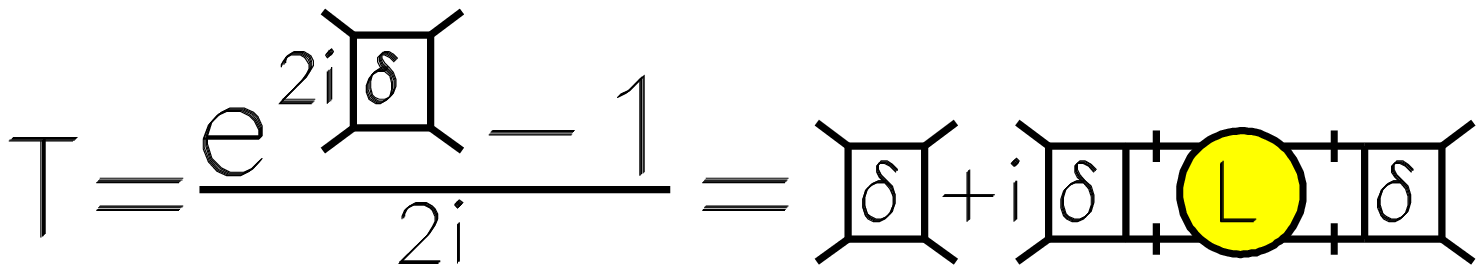}}}}
\vskip 4cm
\end{figure}

Now we can make a very important step. Let us give up some on-shell
conditions,
e.g. let's take $q^2\ne m^2$, $q'^2\ne m^2$ (i.e. two of the interacting
particles are off-shell as for the process $\gamma^* p \rightarrow \gamma^*
p$). Then Eq.(~\ref{eq:expand}) takes the form:
\be
T^{**} =\hat\delta^{**} +i\hat\delta^*\circ L\circ\hat\delta^*
\label{eq:t**}
\ee
where the asterisks mean the number of off-shell momenta. Now we can rewrite
the amplitude in the impact parameter space in the following
way~\cite{petrov}:
\be
T^{**}(s,b)=\delta^{**}(s,b) -
\frac{\delta^*(s,b)\delta^*(s,b)}{\delta(s,b)}+\frac{\delta^*(s,b)\delta^*(s
,b)}{\delta(s,b)\delta(s,b)}T(s,b)
\label{eq:t**1}
\ee
(it is evident that $T(s,\vec b)= T(s,\vert\vec b\vert\equiv b)$ as far as
$\delta(s,\vec b)= \delta(s,\vert\vec b\vert\equiv
b)$Eq.(~\ref{eq:eikonalb}))

The expansion (~\ref{eq:t**}) can be illustrated by the following figure:
\begin{figure}[h]
\vskip -6cm
\hskip -2cm {\vbox to 50mm{\hbox to 50mm{\epsfxsize=120mm
\epsffile{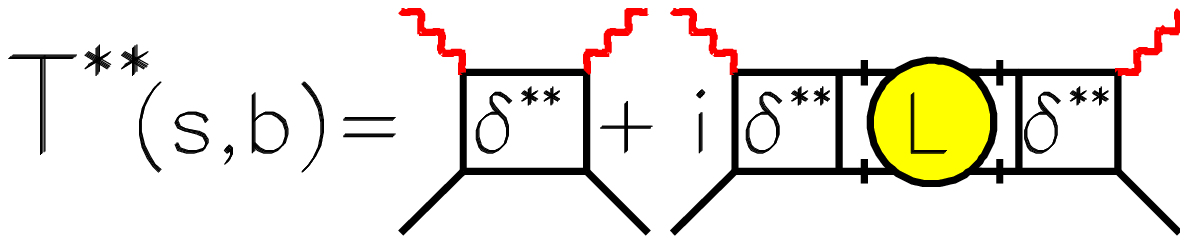}}}}
\vskip 3cm
\end{figure}
 
The case when only one of the particles is off shell can be considered
similarly.
We assume now that $q^2\ne m^2$. Then
Eq.(~\ref{eq:expand}) takes the form

\be
T^{*} =\hat\delta^{*} +i\hat\delta^*\circ L\circ\hat\delta
\label{eq:t*}
\ee

\begin{figure}[h]
\vskip -6cm
\hskip -2cm {\vbox to 50mm{\hbox to 50mm{\epsfxsize=120mm
\epsffile{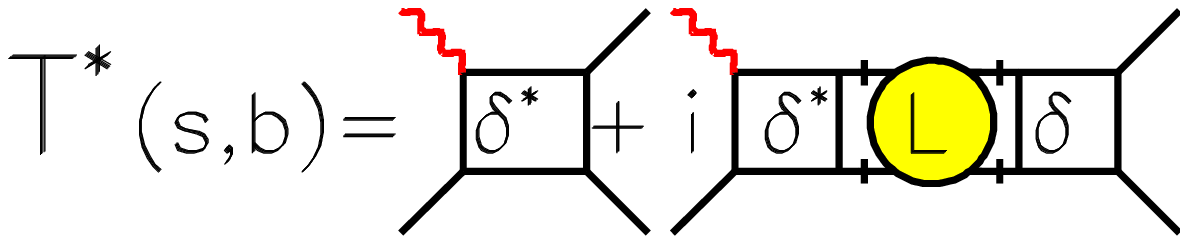}}}}
\vskip 3cm
\end{figure}

or we rewrite it as follows:

\be
T^*(s,b)= \frac{\delta^*(s,b)}{\delta(s,b)}T(s,b)
\label{eq:t*1}
\ee

Now let's choose concrete realizations of the eikonal functions in case of 
virtual particles. On the basis of Eq.(~\ref{eq:eikonalb}) we take the
following parametrizations of off-shell eikonal functions:
\be
\delta^*_{\pm} (s,b)=\xi_{\pm} \frac{c_*(Q^2)}{s_0+Q^2-m^2}
\Big(\frac{s+Q^2-m^2}{s_0+Q^2-m^2}\Big)^{\alpha(0)-1}\frac{e^{-\frac{b^2}{
\rho_*^2}}
}{4\pi
\rho_*^2}
\label{eq:offshelleik*}
\ee
where
\be
\rho_*^2 = 4\alpha'(0) ln\frac{s+Q^2-m^2}{s_0+Q^2-m^2}+r_N^2+r_*^2(Q^2)
\label{eq:offshellrad*}
\ee
and 
\be
\delta^{**}_{\pm} (s,b)=\xi_{\pm} \frac{c_{**}(Q^2)}{s_0+Q^2-m^2}
\Big(\frac{s+Q^2-m^2}{s_0+Q^2-m^2}\Big)^{\alpha(0)-1}\frac{e^{-\frac{b^2}{
\rho_{**}^
2}}}{4
\pi \rho_{**}^2}
\label{eq:offshelleik**}
\ee
where
\be
\rho_{**}^2 = 4\alpha'(0)
ln\frac{s+Q^2-m^2}{s_0+Q^2-m^2}+r_N^2+r_{**}^2(Q^2)
\label{eq:offshellrad**}
\ee
Coefficients $c_*(Q^2)$, $c_{**}(Q^2)$ are supposed to weakly (not in a
powerlike way) depend on $Q^2$. $\xi_{\pm}$ are signature coefficients. 
\noindent
Now let's describe the properties of this model in different kinematic
regions.
\subsubsection{TOTAL CROSS-SECTION}
In accordance with Eq.(~\ref{eq:norm}) we have:
\be
\sigma^{**}_{tot} = \frac{1}{s} ImT^{**}(s,t=0)
\ee
\begin{itemize}
\item {\bf Regge Regime} 
\be
\sigma^{**}_{tot} \rightarrow \frac{(s/Q^2)^\Delta}{Q^2}\Big[c_{**}-
\frac{c_*^2}{c}\Big(\frac{s_0}{Q^2}\Big)^{1+\Delta}\frac{\rho^2}{\rho_*^2}
\Big]
\label{eq:totalregge}
\ee
\item {\bf Bjorken Regime}
\be
\sigma^{**}_{tot} \rightarrow
\frac{c_{**}(Q^2)}{Q^2}\Big(\frac{1}{x}\Big)^\Delta
-\frac{c_*^2}{2c}\cdot \frac{1}{Q^2}\cdot \Big(\frac{1}{x}\Big)^\Delta
\cdot 
\Big(\frac{s_o}{Q^2}\Big)^{1+\Delta} \cdot \frac{ln \frac{Q^2(1-x)}{s_o
x}}{ln\frac{1}{x}}
\label{eq:totalbjorken}
\ee
\end{itemize}
As we see the total cross-section posesses a powerlike behaviour in the
Regge
limit, but this is not a violation of unitarity, as the Froissart-Martin
bound~\cite{fm} cannot be proven for this case and if we put all particles
on the mass shell, then we restore the `normal' logarifmic assymptotical
behaviour $\sigma\sim\ln^2\frac{s}{s_0}$. 
In the Bjorken limit we have strong (powerlike) violation of scaling in the
second term which is not of higher-twist type because of non-integer power. 
\subsubsection{ELASTIC CROSS-SECTION}
For the elastic cross section we have the following
expression (Eq.~\ref{eq:norm}):
\be
\sigma^{*}_{el} = 4 \pi \int_{0}^{\infty} db^2 \Big\vert
\frac{\delta^*}{\delta} T(s,b)\Big\vert^2
\label{eq:elasticcrs}
\ee
As far as $q'^2=\mu^2$, where $\mu$ is the mass of produced particle, it is
natural to set $s_0 = \mu^2$ and now we can derive the following relations:
\begin{itemize}
\item {\bf Regge Regime} 
\be
\sigma^{*}_{el} \rightarrow 16\pi \alpha'(0)\Delta\Big(\frac{c_*}{c}\Big)^2
\Big(\frac{\mu^2}{Q^2}\Big)^{2+2\Delta}(ln\frac{s}{\mu^2})^2
\label{eq:elasticregge}
\ee
\item {\bf Bjorken Regime}
\be
\sigma^{*}_{el} \rightarrow 8 \pi \alpha'(0) \Big(\frac{c_*}{c}\Big)^2
\Big(\frac{\mu^2}{Q^2}\Big)^{2+2\Delta}\frac{(ln(Q^2/x))^2}{ln(1/x)}
\label{eq:elasticbjorken}
\ee
\end{itemize}
As we can easily realize
\be
\frac{\sigma^{*}_{el}}{\sigma^{**}_{tot}} \rightarrow 0
\label{eq:eltot}
\ee
Now we are ready to turn to describing the proton structure function
$F_2^p(x,Q^2)$
\section{THE MODEL FOR $F_2^p(x,Q^2)$}
The proton structure function $F_2^p(x,Q^2)$ can be connected to the
transverse cross-section $\sigma^{**}_T(W,Q^2)$ of the $\gamma^* + p
\rightarrow X$ process by the following relation:
\be
\sigma^{**}_T(W,Q^2)= \frac{4\pi ^2
\alpha}{Q^2(1-x)}\frac{1+\frac{4m_p^2x^2}{Q^2}}{1+R(x,Q^2)}F_2^p(x,Q^2)
\label{F2}
\ee
where $W^2=\frac{Q^2}{x}-Q^2+m_p^2$,
$R(x,Q^2)=\frac{\sigma^{**}_L}{\sigma^{**}_T}$. As far as $R(x,Q^2)$ is
deemed to be small, we let it be equal to $0$ in what follows, i.e. we
suppose that the total coss-section is the same as the transverse one.

In the following consideration we restrict ourselves within small
$x$ region ($x<10^{-2}$) so that we would be able to use the asymptotic
formula (~\ref{eq:totalregge}) which explicitly gives us the effects of
unitarization in our model.
Using Eq.~\ref{eq:offshelleik**}and Eq.~\ref{eq:offshellrad*},
Eq.~\ref{eq:totalregge} can be rewritten as ($s\equiv W^2$) 
\be
\begin{array}{r}
\sigma^{**}_{tot} \rightarrow
\frac{((W^2+Q^2-m_p^2)/(W_0^2+Q^2-m_p^2))^\Delta_{
P}}{(W_0^2+Q^2-m_p^2)}\cdot \\
\cdot\Big[c_{
**}(Q^2)-
\frac{c_*^2(Q^2)}{c}\Big(\frac{W_0^2-\mu^2-m_p^2}{W_0^2+Q^2-m_p^2}\Big)^{1+
\Delta_{P}
}\frac{\rho^2}{\rho_*^2}
\Big] \\
\end{array}
\label{eq:totalregge1}
\ee
To derive this formula we have made the following assumptions:
\begin{itemize}
\item We suppose that the amplitude for $\gamma^* p$ scattering is
proportional~\cite{vectordominance} to the amplitude of virtual vector meson
scattered on $p$ and
this 
``effective" vector meson is $\rho_0$ with mass $\mu=0.77$ (GeV), i.e.
\be
T_{\gamma^* p\rightarrow \gamma^* p}(W,Q^2,t)=k\cdot T_{\rho_0^*
p\rightarrow
\rho_0^* p}(W,Q^2,t)
\ee
where $k$ is some constant.
\item As far as we are using the asymptotic formulas, we neglect the real
part of the signature coefficient for pomeron (as far as this is
proportional to $\Delta_{P}\simeq 0.1$) and set the signature
coefficient be
$i$
\item The parametrizations of $c_{**}$ and $c_*$ are following:
\be
\begin{array}{l}
c_{**}(Q^2)= c^{**}\\
c_*(Q^2)= c^*+c^*_1ln(\frac{Q_0^2+Q^2}{Q_0^2})^3\\
c = c_*(-\mu^2)\\
\end{array}
\ee
where $Q_0^2=1.0\;(GeV^2)$ and $c^{**},\; c^{*},\; c^{*}_1 $ are
parameters.
\item The radii $\rho^2,\;\rho_*^2$ are parametrizied as follows:
\be
\begin{array}{l}
\rho_*^2(W,Q^2)=4\alpha'(0)ln\frac{W^2+Q^2-m_p^2}{W_0^2+Q^2-m_p^2}+r^2/(Q_0^
2
+Q^2) \\
\rho^2(W)= \rho_*^2(W,-\mu^2) \\
\end{array}
\ee
where $r$ is a parameter.
\end{itemize}
Eventually we have the following expression for $F_2^p(x,Q^2)$:
\be
\begin{array}{l}
F_2^p(x,Q^2) =
\frac{1}{4\pi^2\alpha}\frac{Q^2(1-x)}{1+\frac{4m_p^2x^2}{Q^2}}\cdot \\
\cdot
\frac{((W^2+Q^2-m_p^2)/(W_0^2+Q^2-m_p^2))^\Delta}{(W_0^2+Q^2-m_p^2)}\cdot \\
\cdot\Big[c_{
**}(Q^2)-
\frac{c_*^2(Q^2)}{c}\Big(\frac{W_0^2-\mu^2-m_p^2}{W_0^2+Q^2-m_p^2}\Big)^{1+
\Delta
}\frac{\rho^2}{\rho_*^2}
\Big] \\
\end{array}
\label{eq:myf2}
\ee
\section{RESULTS}
As we have said, for the fit we used the data with $x<10^{-2}$ and thus,
from the whole set of 1265 data we extracted 401 experimental points. Having
used 5 free parameters we achieved $\chi^2=0.914$. The fitted parameters 
are given in Table~\ref{tab:1} (The intercept and the slope of the
pomeron is derived from the fit of nucleon-nucleon cross-sections
in~\cite{Prokudin}).

\begin{figure}[htb]
\centering
\hskip 0cm {\vbox to 100mm{\hbox to 100mm{\epsfxsize=100mm
\epsffile{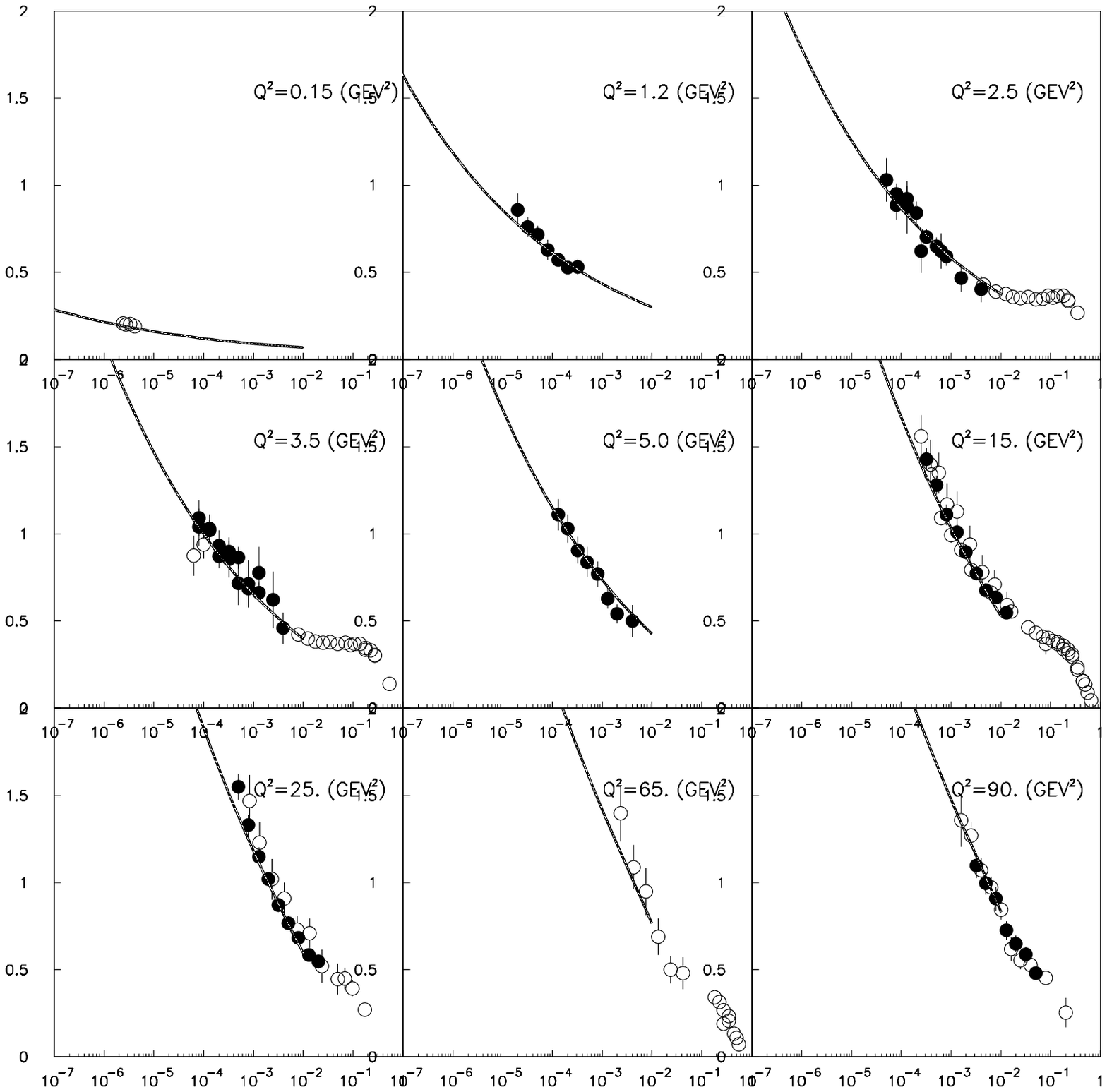}}}}
\vskip 1.5cm
\caption{Experimental data for the proton structure function $F_2^p(x,Q^2)$
at low and intermediate $Q^2$ and predictions of the model.}
\label{fig:f2}
\end{figure}

\begin{figure}[htb]
\centering
\hskip 0cm {\vbox to 100mm{\hbox to 100mm{\epsfxsize=100mm
\epsffile{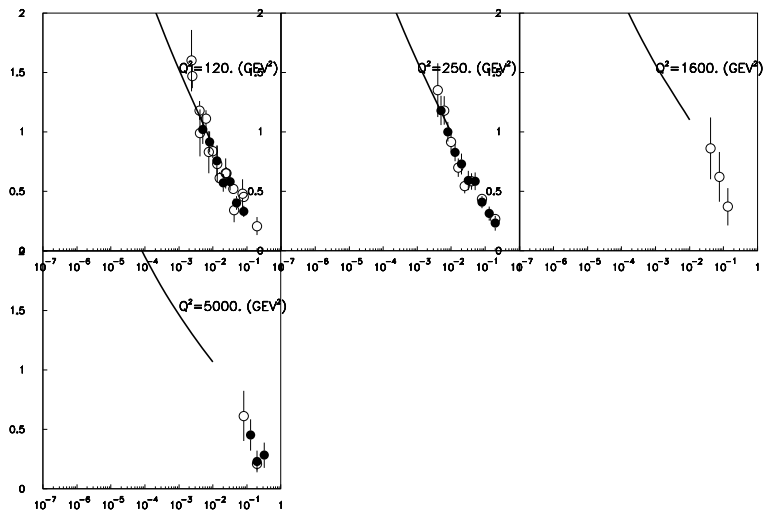}}}}
\vskip -2cm
\caption{Experimental data for the proton structure function $F_2^p(x,Q^2)$
at intermediate and high $Q^2$ and predictions of the model.}
\label{fig:f21}
\end{figure}

\begin{table}[h]
\caption{Parameters obtained by fitting to the data. \label{tab:1}}
\begin{center}
\begin{tabular}{|c|c|c|c|}
\hline
$\Delta_{P}$(fixed) &  0.11578 &$\alpha'_{P}$(fixed)& 0.27691\\
$c^{**}$&  7.5756 & $c^{*}$& 3.0036 \\
$c_1^*$ & 0.030931 & $r^2$ & 117.89 \\
 $Q_0^2(GeV^2)$(fixed)& 1.0 &$W_0(GeV)$ & 1.6336 \\
\hline
\end{tabular}
\end{center}
\end{table}
The results of fitting are presented in Figures (~\ref{fig:f2}) and
(~\ref{fig:f21})

\subsection{$x$-SLOPE OR ${\partial lnF_2^p(x,Q^2)/\partial ln(1/x)}$}
\begin{figure}[htb]
\centering
\epsfig{file=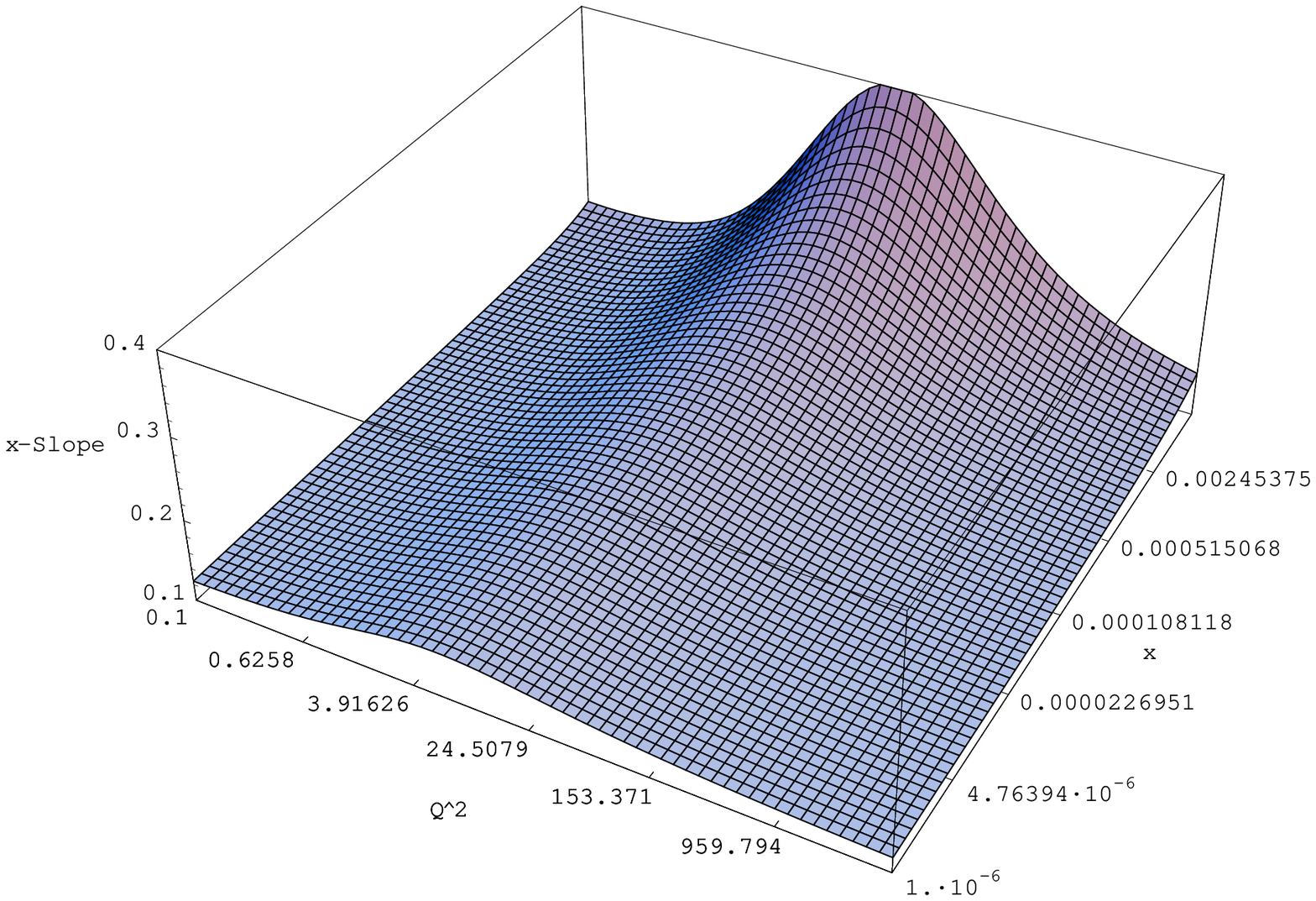, height=90mm}
\caption{The $x$-slope $\frac{\partial lnF_2^p(x,Q^2)}{\partial ln(1/x)}$ 
as a function of $x$ and $Q^2$.}
\label{fig:x}
\end{figure}

The data on $F_2^2(x,Q^2)$ show a tendency of fast growth with decrease of
x. This is so-called HERA effect. Our model suggests that this effect will
ease down with increase of $Q^2$. This can be seen in Figure~\ref{fig:x}.
The same effect is predicted in the Dipole pomeron model~\cite{Martynov}. We
believe that new experimental observations in the kinematic region of
$100\le Q^2\le 1000$ and $x\le 10^{-2}$ will help to verify
existence of this phenomenon.

\begin{figure}[htb]
\centering
\vskip -6cm
\hskip 0cm {\vbox to 60mm{\hbox to 60mm{\epsfxsize=60mm
\epsffile{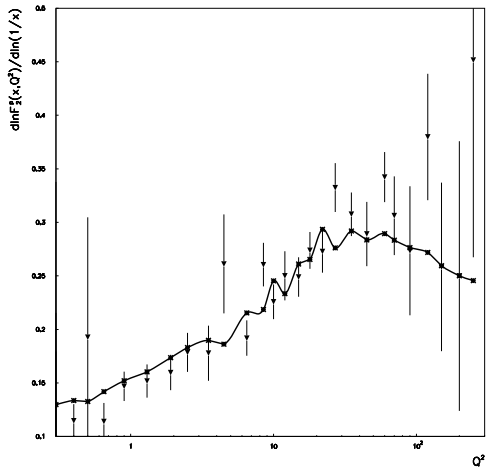}}}}
\vskip 1cm
\caption{The effective intercept measured experimentally and comparison with
our model. (The asterisks denote the points where $\frac{\partial
lnF_2^p(x,Q^2)}{\partial ln(1/x)}$ was calculated.)}
\label{fig:xexp}
\end{figure}

The effective intercept, which is mesaured by experimentalists assuming that
$F_2^p \propto (1/x)^{\Delta_{eff}(Q^2)}$ can be
approximately identified with the $x$-slope if the slope depends weakly on
$x$. We performed calculation of the slope and the comparison with the
experimental data is in the Figure~\ref{fig:xexp}. As is seen the
experimental results are well described by the model. 

\subsection{$Q$-SLOPE OR ${\partial F_2^p(x,Q^2)/\partial ln(Q^2)}$}
\begin{figure}[htb]
\centering
\hskip 0cm {\vbox to 60mm{\hbox to 60mm{\epsfxsize=60mm
\epsffile{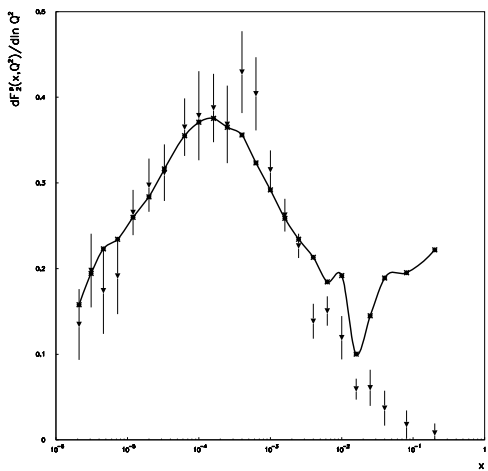}}}}
\vskip 1cm
\caption{The $Q$-slope $\frac{\partial F_2^p(x,Q^2)}{\partial ln(Q^2)}$ (The
asterisks
denote the points where $\frac{\partial lnF_2^p(x,Q^2)}{\partial ln(1/x)}$
was calculated.)}
\label{fig:q}
\end{figure}

\begin{figure}[htb]
\centering
\epsfig{file=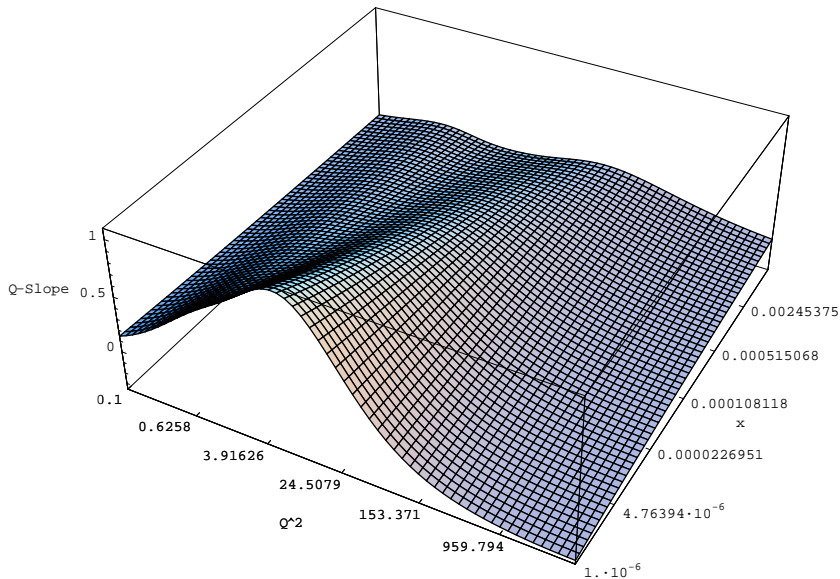, height=90mm}
\caption{The $Q$-slope $\frac{\partial F_2^p(x,Q^2)}{\partial ln(Q^2)}$ 
as a function of $x$ and $Q^2$.}
\label{fig:qexp}
\end{figure}

The data on the $Q$-slope shows a peak at $Q^2 \sim 1-5\;GeV^2$. This peak
is interpreted as a transition from Regge behaviour to a perturbative QCD
regime. We performed calculation in the framework of our model and the
result is in Figure.~\ref{fig:q}. As is seen, there is no contradiction in 
the behaviour of the $Q$-slope and Regge regime. We also want to emphasise
that the determination of the transition region depends on the path on the
two-dimentional surface of ${\partial F_2^p(x,Q^2)/\partial ln(Q^2)}$ and it
may lead to the $x$-dependence of the position of the peak~\cite{Martynov}.
To demonstrate this we show this surface calculated in our model in
Figure~\ref{fig:qexp}.

\section{CONCLUSION}

The proton structure function $F_2^p(x, Q^2)$ is described in the framework
of extended Regge-eikonal approach. It is argued that for this one does not
need additional Regge poles with intercepts depending on $Q^2$. The
experimental data on $x$- and $Q$-slope are also described fairly well. The
model predicts damping of the HERA effect (the similar prediction is made in
the
framework of the Dipole pomeron model~\cite{Martynov}).   

\section*{ACKNOWLEDGMENTS}
We would like to thank E. Martynov and A. De Roeck for providing us with
experimental data 
and we are grateful to E. Martynov for valuable discussions. One of us
(A.P.) is indebted to ICTP HEP
division authorities for a kind invitation and hospitality during his visit
to ICTP, where a part of this work was done.

\newpage

\end{document}